\documentclass[ a4paper, 12pt]{article}
\usepackage[text={16cm,24cm},centering]{geometry} % geometrija stranice
\usepackage[utf8]{inputenc}
\usepackage{amssymb ,amsmath, amsthm}
\usepackage{graphicx} % paket za umetanje slika
\usepackage{ulem}
\usepackage{multirow}
\usepackage{authblk}

\usepackage{color}
\usepackage{float}

\usepackage[font=footnotesize,labelfont=bf]{caption}
\usepackage{enumerate}
\newcounter{Section}
\newcommand{\Keywords}[1]{\par\noindent
{\small{\em Keywords\/}: #1}}

\title{High Frequency Trading and Mini Flash Crashes
\footnote{The authors gratefully acknowledge funding from the European Community's Seventh Framework Programme FP7-PEOPLE-ITN-2008 under grant agreement number PITN-GA-2009-237984. }}
\author{Anton Golub}
\affil[1]{Olsen Ltd.}
\author{John Keane}
\affil[2]{University of Manchester - School of Computer Science}
\author{Ser-Huang Poon}
\affil[3]{Manchester Business School}

\begin{document}
\maketitle
\abstract{We analyse all Mini Flash Crashes (or {\it Flash Equity Failures}) in the US equity markets in the four most volatile months during 2006-2011. In contrast to previous studies, we find that Mini Flash Crashes are the result of regulation framework and market fragmentation, in particular due to the aggressive use of Intermarket Sweep Orders and Regulation NMS protecting only Top of the Book. We find strong evidence that Mini Flash Crashes have an adverse impact on market liquidity and are associated with Fleeting Liquidity.\\

}
\newpage 

\section{Introduction}
Much attention in recent years has focused on traders who reportedly account for upwards of 50 percent of market volume and whereby the ability to make consistent profits largely depends on speed. The so called ``race to zero'' has led to the rise of an entire technology sub-industry. For example, the \$300 million Project Express by Hibernia Atlantic aims to reduce latency across the Atlantic Ocean by 5 milliseconds,\footnote{1 millisecond is one thousandth of a second; 1 microsecond is one millionth of a second.} which will make it the world’s fastest transatlantic cable when it opens in 2013 and the first to achieve round-trip trading speeds of less than 60 milliseconds.\footnote{``High-Speed Trading: My Laser Is Faster Than Your Laser'', Philips, M. \textit{BloombergBusinessweek: Markets \& Finance}, April 23, 2012.} Exchanges also play their part in this ``arms race'': the latest addition is the Singapore Exchange's (SGX) \$250 million worth Reach Initiative, which claims to have the fastest trading engine, with average order response time of less than 90 microseconds door-to-door - the fastest order execution ever tested.\\

High frequency trading (HFT) gained prominence in the media after May 6th 2010, the day when the U.S stock market experienced one of its most severe price drops in history: the Dow Jones Industrial Average (DJIA) index dropped by almost 9\% from the open - the second largest intraday point swing, 1,010.14 points, and the biggest one-day point decline of 998.5 points. The crash was so rapid that the index tumbled 900 points in less than 5 minutes, but recovered the bulk of its losses in the next 15 minutes of trading. This event is now known as the ``May 6th 2010 Flash Crash''. Such a large swing raised much concerns about the stability of capital markets, resulting in a US Securities and Exchange Commission (SEC) investigation of the US equity markets (SEC, 2010). This report claimed that the crash was ignited by a sell algorithm of a large mutual fund\footnote{ The mutual fund is alleged to be Waddell \& Reed Financial Inc.} executing a \$4.1 billion sell trade in the eMini S{\&}P500 futures. While HFT did not spark the crash, it does appear to have created a ``hot potato'' trading volume that contributed to the crash. In contrast, Nanex (Nanex, 2010) concluded that quote stuffing and delays in data feeds created a positive feedback loop that set-off the crash. Furthermore, Nanex found that the mentioned sell algorithm in fact did not have a market impact as it has sold on the offer, thereby providing (upside) liquidity to the market.\\

In this paper we study \textit{Mini Flash Crashes} in US equity markets.  Mini Flash Crashes are abrupt and severe price changes that occur in an extremely short period. They are scaled down versions of the May 6th 2010 Flash Crash. Figure 1 shows an example of a \textit{down Mini Flash Crash} that occurred on October 7th 2008 to the stock price of Goldman Sachs Group, Inc. (NYSE ticker: GS) – when a 1.6\% price drop occurred in less than 400 milliseconds. The price dropped \$1.89 and then immediately returned to its previous levels. Mini Flash Crashes only became known to the general public in late 2010 when Nanex preformed a thorough analysis of the US stock markets in the aftermath of the May 6th Flash Crash, attributing these crashes to HFT activity (Nanex, 2010). Perhaps the most well known recent Mini Flash Crash is that of the BATS Initial Public Offering (IPO).\footnote{``Bats: The Epic Fail of the Worst IPO Ever'', Beucke, D., \textit{BloombergBusinessweek: Markets \& Finance}, March 23, 2012.}\footnote{``BATS Flash Crash: Here's What Happened'', Albinus, P., \textit{Advanced Trading: Exchanges/ECNs}, March 26, 2012.} Shares of BATS, the third largest  US electronic exchange, started trading on March 23rd, 2012 at 11:14:18.436 with an initial price of \$15.25. Within 1.5 seconds, the price dropped to \$0.0002, as a series of Intermarket Sweep Order (ISO) were executed on NASDAQ. The fate of the IPO was sealed as the shares were indefinitely withdrawn from trading later that day.\\

\begin{figure}[h]
    \begin{center}
        \includegraphics[scale=0.5]{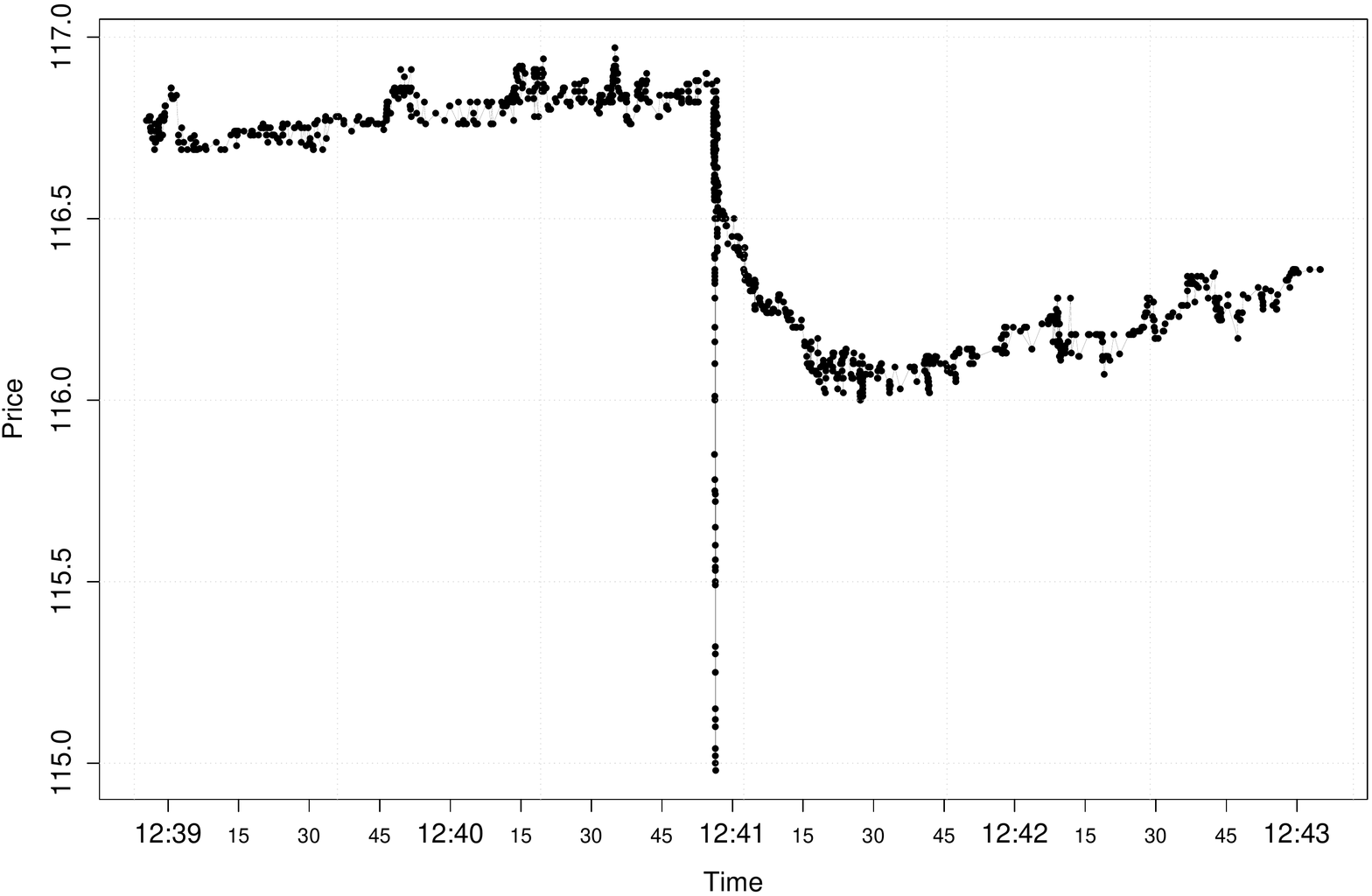}
        \caption{\footnotesize \textbf{Example of a Mini Flash Crash.} The figure shows a Mini Flash Crash occurring at NYSE in the stock of Goldman Sachs Group, Inc. (NYSE ticks: GS) on October 7th 2008, resulting in a 1.6\% price drop in less than 400 milliseconds. The crash consisted of 58 trades, where the last 57 were marked ISO, for a total volume of 86000 shares, reducing the price \$1.89 from its initial values.}
    \end{center}
\end{figure}

This paper investigates and analyses the causes of Mini Flash Crashes. Whilst Mini Flash Crashes have been high profile in the financial community, there has not been any attempt to explain these crashes. In Johnson et. al. (2012), the authors suggest that the crashes may be a result of interaction between several trading algorithms, or a positive feedback loop induced by market environment. Moreover, they argue that: \textit{``...these ultrafast black swans are not simply the product of some pathological regulatory rule for crashes...''}. Contrary to Johnson et. al. (2012), we find evidence that Mini Flash Crashes are the product of market regulation. Specifically, we conclude that the crashes predominantly happen due to the aggressive use of ISO and the fact that Regulation National Market System (NMS) only protects the Top of the Book. In practice this manifest as a burst of trades from one exchange, as remaining orders rips through the book due to the lack of Depth of the Book protection. We characterise two types of Mini Flash Crashes: \textit{ISO-initiated} and \textit{auto-routing-initiated}. We find that Mini Flash Crashes are mostly ISO-initiated and have adverse impact on liquidity conditions resulting in wider quotes, increased number of locked or crossed quotes, and reduction of quoted volume. Further, we find Fleeting Liquidity to be related with Mini Flash Crashes. The logit regression results show that Fleeting Liquidity is associated with ultra-fast Mini Flash Crashes, of small volume and a small number of trades.\\

The rest of this paper is organized as follows; Section 2 provides a survey of previous literature and defines Mini Flash Crashes. The dataset used in the research is described in Section 3. Section 4 presents statistics on Mini Flash Crashes, and the definition of two types of crashes. Section 5 examines the impact of Mini Flash Crashes on market liquidity, and relates \textit{Fleeting Liquidity} to crash characteristics. Finally, Section 6 concludes and discusses the weaknesses of current regulation and suggests ways to restrain Mini Flash Crashes.
\section{Background}
This work belongs to a growing field of research on HFT and market activity in millisecond environments. Previous research has mainly focused on the cause of the May 6th 2010 Flash Crash. The joint report of the SEC and the Commodity Futures Trading Commission  (2010) attributed a sell order in the eMini S\&{P} 500 futures market as the cause of the Flash Crash. While HFTs did not start the crash they generated a ``hot-potato'' volume accelerating mutual fund's sell side pressure and contributing to the sharp price declines that day. On the other hand, Nanex (Nanex, 2010) found that quote saturation and NYSE Consolidated Quotation System (CQS) delays, combined with negative news from Greece and sale of eMini S\&{P} 500 futures, ”...was the beginning of the freak sell-off which became known as the flash crash”. Nanex rejected claims the crash was due to NYSE ``Slow Quote'' mode or Liquidity Replenishment Point (LRP)\footnote{NYSE implements price-bands known as ``Liquidity Replenishment Points" (LRP) intended to dampen volatility in a given stock by temporarily converting from an automated market to a manual auction market when a price movement of sufficient size is reached. In such a case, trading on NYSE in that stock will go to ``Slow Quote'' Mode and automatic executions will cease for a time period ranging from a fraction of a second to a minute or two to allow the Designated Market Maker to solicit and/or contribute additional liquidity before returning to an automated market. LRP limits vary according to each security’s share price and average daily volume within specified ranges, generally falling between 1\% and 5\% of share price.}, and movements in the price of Apple Inc. (AAPL) and stub quotes\footnote{Stub quotes are order placed well off a stock's market price.}. Further, Nanex analysed the mutual fund's trade executions in the June 2010 eMini futures contract on May 6th 2010 and showed that the algorithm neither took nor required liquidity at any stage, and that the bulk of the sell trades occurred after the market bottomed and was on the recovery trend.  Kirilenko et. al. (2011) use audit-trail data and examined the trades in the eMini S\&{P} 500 stock index futures market on May 6 and also conclude that HFTs did not trigger the Flash Crash, though their responses caused a ``hot potato'' effect on the unusually large selling pressure on that day exacerbated market volatility.\\

Easly, Lopez de Prado, O'Hara (2011) show the Volume-Synchronized Probability of Informed Trading (VPIN) captures the increasing toxicity of the order flow prior to the May 6th 2010 Flash Crash. Furthermore, Wood, Upson and McInish (2012) find that trading aggressiveness was significantly higher for that entire day, proxied by the use of ISOs\footnote{An ISO is a limit order designated for quick and automatic execution in a specific market center and can be executed at the target market even when another market center is publishing a better quotation. Without the need to search for the best quotation during their execution, such orders can possibly achieve faster execution than regular market orders, even though both types of orders demand liquidity and move price. When submitting an ISO, the submitting investor also needs to fulfill Regulation NMS order-protection obligations by concurrently sending orders to other market centres with Protected Quotations. Each order in the package must be marked as an ISO and the orders sent to the market(s) posting Protected Quotation must be of sufficient quantity to match their displayed depth at the Top of the Book. Importantly, ISO orders are not subject to auto-routing and are identified with a trade indicator "F" by the trade initiator. The sweep ends when the order quantity is satisfied, the limit price is reached, or when the order hits an intervening LRP set by the market. Although trade-through can, and does, occur, ISOs are deemed to satisfy the Order Protection Rule under Regulation NMS. Chakravarty, Jain, Upson, and Wood (2009) provide detailed regulatory background on ISOs.}. They show that the information content of ISO trades on the day of the May 6th Flash Crash was highly informed. During the May 6th Flash Crash, while the ISO volume was at 32\% of total volume, ISO trades accounted for over 50\% of the contribution to the price variance. They show that as liquidity decreases in the market, it leads to an increase in ISO use, accelerating market decline whilst speeding-up the market recovery - they recommend considerations of an ISO halt during periods of high market wide volatility. Cespa and Foucault (2012) argue that the May 6th 2010 Flash Crash is due to the self-reinforcing relationship between price informativeness and liquidity leading to contagion and fragility in the market. Finally, Paddrik et. al. (2011), create a zero-intelligence agent based model of eMini S\&{P} 500 futures market and validate the simulated market against empirically observed characteristics of price returns and volatility. They illustrate the applicability of the simulation and present experimental results which examine the hypothesis for the cause of the May 6th 2010 Flash Crash. Biais, Foucault, Moinas (2012) claim that since HFTs can process information on stock values faster than slow traders, adverse selection could lead to systemic risk events, such as the May 6th 2010 Flash Crash.\\

Unlike all previous studies described above, we examine the characteristics of Mini Flash Crashes. Mini Flash Crashes, or Flash Equity Failures, were first identified by Nanex Llc., therefore we adopt their definition (Nanex, 2010). To qualify as a \textbf{\textit{down crash}} candidate, the stock price change has to satisfy the following conditions:
\begin{enumerate}[(i)]
    \item it has to tick \textit{down} at least 10 times before ticking \textit{up},
    \item price changes have to occur within 1.5 seconds,
    \item price change has to exceed -0.8\%.
\end{enumerate}
Likewise, to qualify as an \textbf{\textit{up crash}} candidate, the stock price change has to satisfy the following conditions:
\begin{enumerate}[(i)]
    \item it has to tick \textit{up} at least 10 times before ticking \textit{down},
    \item price changes have to occur within 1.5 seconds,
    \item price change has to exceed 0.8\%.
\end{enumerate}

As far as we are aware, the only comparable work on Mini Flash Crashes is that by Johnson et. al. (2012), in which a set of 18,520 Mini Flash Crashes occurring between 2006 and 2011 are analysed. Johnson et. al. (2012) provide empirical evidence for, and an accompanying theory of, an abrupt system-wide transition from a mixed human-machine phase to a new all-machine phase characterized by frequent black swan events of ultrafast duration. This theory quantifies the systemic fluctuations in these two distinct phases in terms of the diversity of the system’s internal ecology and the amount of global information being processed.  The ten most
susceptible entities are found to be major international banks, suggesting that there is a hidden relationship between these ultrafast ``fractures'' and the slow ``breaking'' of the global financial system post-2006. Further, they suggest that: \textit{``...these ultrafast black swans are not simply the product of some pathological regulatory rule for crashes...''}. In contrast, we find evidence  that regulatory framework and market fragmentation are the primary cause of Mini Flash Crashes. 

\section{Dataset}
The dataset used in this study includes all quotes disseminated by the Consolidated Quotation System (CQS) and trades disseminated by Consolidated Tape System (CTS) for NYSE, NYSE Amex, NYSE Arca listed securities, and Unlisted Trading Privileges (UTP) Quote Data Feed disseminated quotes and UTP Trade Data Feed disseminated trades for NASDAQ listed securities, as well as trades reported from Alternative Trade Facilities (ADF) and Trade Reporting Facilities (TRF)\footnote{Trades that are executed off the exchange, including dark pool trades, internalized trades and ECN trades, are reported to the consolidated tape indirectly through ADF and TRF. These trades are absent from our analysis as they are executed at the NBBO.}. The messages are placed in the database as they are aggregated from the Security Information Processor (SIP), and correspond to the information that professional traders receive in real-time. Each message is time stamped to a precision of 25 milliseconds, and contains information relating to time, price, size, exchange issuance and eventual trade or quote condition. \\

In this study we have analyzed four months of Mini Flash Crashes occurring in September to November 2008, and May 2010, giving a total of 5140 crashes, made up of 2760 up crashes and 2380 down crashes. Figure 2. presents the daily number of up and down crashes from the period 2006-2011. The shadowed area indicates the crashes analysed in this study. The purpose of studying Mini Flash Crashes in the targeted four months is twofold: (i) the corresponding periods were highly volatile including the unfolding of the global financial crisis, and the collapse of Lehman Brothers and the May 6th 2010 Flash Crash; (ii) analysis of Mini Flash Crashes means processing billions of trades and quotes, therefore we selected months with the most crashes - hence our dataset  covers over one-quarter (27.75\%) of all Mini Flash Crashes reported to have occurred in the period of 2006-2011. \\

\begin{figure}[H]
	\footnotesize
    \begin{center}
        \includegraphics[width=6in,height=5.5in]{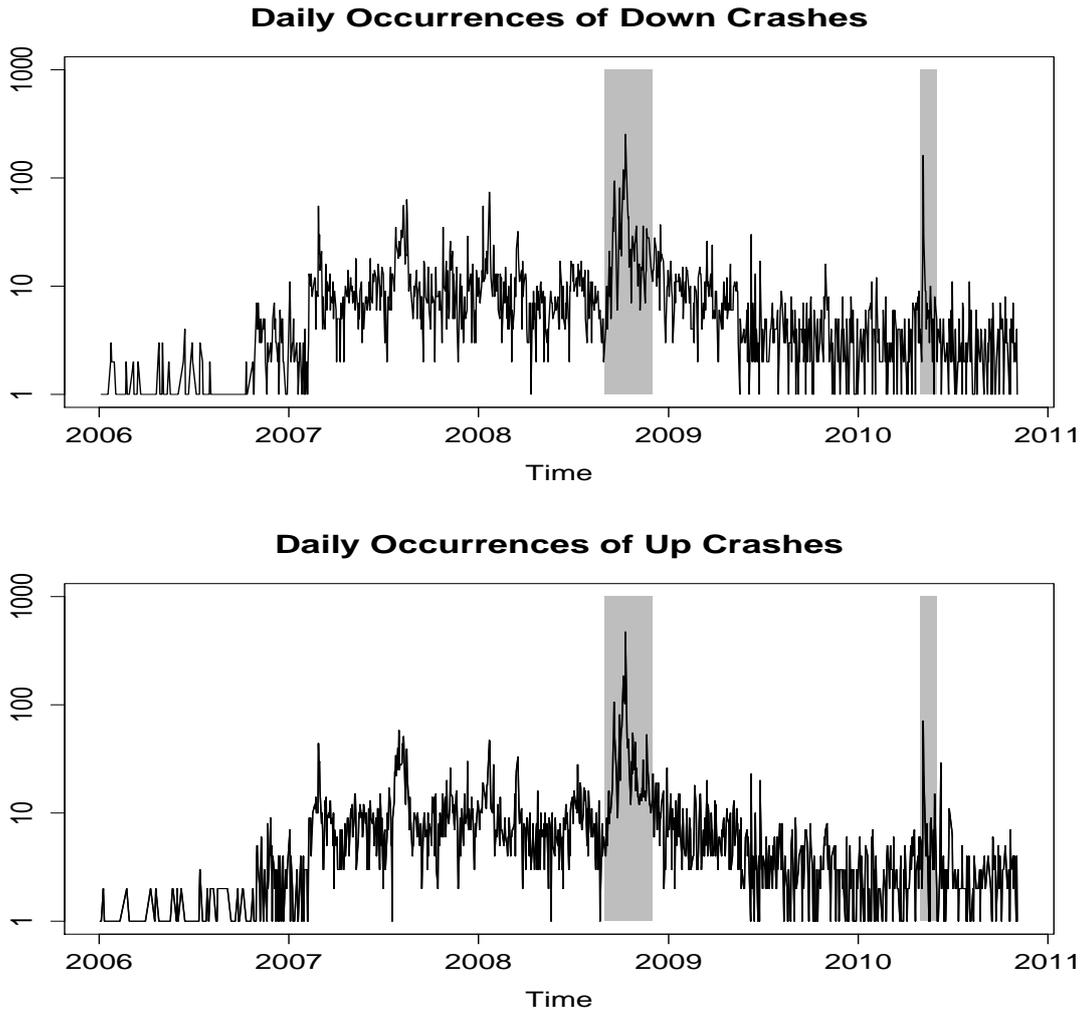}
        \caption{\footnotesize \textbf{Daily occurrences of Up and Down Crashes 2006-2011}. The shadowed area marks the crashes analyzed in this study, corresponding to months September to November 2008, and May 2010. The Y axis presents the number of daily occurrences and is on a log scale.}
    \end{center}
\end{figure}

\section{Results and Analyses}
In this section we analyse the characteristics and mechanics of two types of Mini Flash Crashes, ISO-initiated and auto-routing initiated outlined later in this section. Based on this two definitions, we have identified 71.49\% of the Mini Flash Crashes in our dataset, and conclude that the overwhelming majority of crashes is due to aggressive use of ISOs.\\

We begin our analysis by presenting the descriptive statistics of Mini Flash Crashes. Table 1. reports the average percentage price change in a crash, the average duration of the crash, the average trade size in a crash, the average number of trades per crash and the average number of ISO trades as a percentage of total number trades. When computing the average percentage price change in a crash, we have excluded crashes in penny-stocks, as well as percentage price changes exceeding 100\% as they heavily bias the average value. As the crashes occur on a single exchange, we report the percentages of crashes at six exchanges, with NYSE, Nasdaq and Arca covering the majority of the crashes. Further, an extremely high number of ISO marked trades can be immediately noticed, where they constitute the overwhelming majority of trades in the crashes. This is an evidence on excessive use of ISOs corresponds to the findings of Chakravarty, Jain, Upson, and Wood (2009) who attribute their findings to informed investors. Here we cannot attribute high ISO usage to informed trading, as the trades in the crashes are executed well off market prices. \\

\begin{center}
\begin{table}[H]
	\captionsetup{singlelinecheck=off}
	\footnotesize
	\caption[try out]{\footnotesize \textbf{Descriptive Statistics and Classification of Mini Flash Crashes}. \\
	This table shows descriptive statistics for the Mini Flash Crashes, as well as monthly descriptive statistics for September to November 2008, and May 2010:
	\begin{enumerate}[(i)]
	\item \textit{Avg \% Change} is the average total percentage price change in a crash. We have excluded Mini Flash Crashes occurring in penny-stocks and those Mini Flash Crashes with percentage price changes exceeding 100\%, as they heavily bias the average value; 
	\item \textit{Avg Time} is the average time of the crash in milliseconds;
	\item \textit{Avg Trade Vol} is the average total volume of the trades;	
	\item \textit{Avg No of Trades} is the average number of trades; 
	\item \textit{ISO Trades} is the average number of ISO marked trades, as a percentage of total number of trades. 
	\end{enumerate}}
	\begin{center}
	\begin{tabular}{l l l l l l}
	& Sep 2008 & Oct 2008 & Nov 2008 & May 2010 & Total \\ \hline
	Total Crashes & 1058 & 2919	& 672 & 	491 & 5140 \\
	~~~~Up Crashes & 527	& 1709 &	 337 & 187 & 2760 \\
	~~~~Down Crashes & 531 & 1210 & 335 & 304 & 2380 \\
	& & & & & \\
	ISO initiated & 719 (67,96\%) &	1922 (65,84\%)	& 519 (77,23\%) & 328 (66,80 \%) & 3488 (67,85\%) \\
	auto-routing initiated & 29 (2,74\%)	& 171 (5,86\%) & 11 (1,63\%) & 27 (4,02\%) & 238 (4,64\%) \\
	unclassified & 310 (29,30\%) & 826 (28,30\%)	& 142 (21,13\%)	& 136 (27,70\%) & 1414 (27,51\%) \\ \hline
	& & & & & \\
	Avg \% Change & 2,69\% & 2,14\% & 2,17\% & 4,12\% & 2,45\% \\ 
	Avg Time &	437 & 432 &	374 & 319 &	413 \\
	Avg Trade Vol & 12465,97 & 13900,13	& 18390,18 & 14810,52 & 14352,39 \\ 
	Avg No of Trades & 21,15	& 17,98 	& 19,08	 & 19,72 & 18,94 \\ 
	~~~~ISO Trades & 89,53\% & 85,80\% &	89,95\%	 & 84,30\% & 86,98\% \\ 
	
	& & & & & \\
	by Exchanges & & & & & \\ 
	~~~~NYSE	 & 37,09\% & 67,56\% & 58,14\% & 58,41\% & 59,29\% \\ 
	~~~~NASDAQ	& 32,58\% & 18,75\% & 23,94\% & 19,39\% & 22,28\% \\ 
	~~~~ARCA & 29,03\% & 13,15\%	 & 17,43\% & 19,39\% & 17,52\% \\ 
	~~~~AMEX & 1,30\% & 0,46\% & 0,16\% & 1,17\% & 0,65\% \\ 
	~~~~BATS & 0,00\% & 0,00\% & 0,33\% & 1,40\%	& 0,19\% \\ 
	~~~~ISE & 0,00\% & 0,08\% & 0,00\% & 0,23\% & 0,07\% \\ 
	\end{tabular}
	\end{center}
\end{table}
\end{center}

\newpage

In order for a crash to be classified as \textbf{\textit{ISO-initiated}}, it has to satisfy the following:
\begin{enumerate}[(i)]
	\item the conditions defining a Mini Flash Crash set out in Section 2 must be satisfied,
	\item the trades constituting the Mini Flash Crash have to be marked ISO, except for the first $k$ trades that can be marked as non-ISO if they are executed within the least aggressive market-wide available best quotes\footnote{We require that these quotes are not \textit{stub quotes}.} in the previous one (1) second. 
\end{enumerate}
The motivation for allowing the first $k$ trades to be marked as non-ISO is threefold. Firstly, the $k$ mentioned trades can be part of the Order Protection compliance of ISO trades initiated at another exchange, followed by regular marked trades at the domestic exchange, and therefore not a part of the ISO package of trades constituting a Mini Flash Crash that sweeps the order book. 
Secondly, the first $k$ trades that can be non-ISO, may be a part of an auto-routing procedure of another exchange complying with the Order Protection Rule, and therefore not relevant to the mechanics of ISO-initiated Mini Flash Crashes, as the auto-routing procedure will only send the part of the order matching the displayed depth at the Top of the Book, while the remaining balance of the order will be filled elsewhere. 
Thirdly, allowing the eventual non-ISO trades to be executed within the least aggressive quotes in the previous one (1) second allows the reference prices for execution to be compliant with Flicker Quote Protection.\\

Figure 3 illustrates the mechanics of ISO-initiated Mini Flash Crashes.\\

\begin{figure}[h]
    \begin{center}
        \includegraphics[scale=0.45]{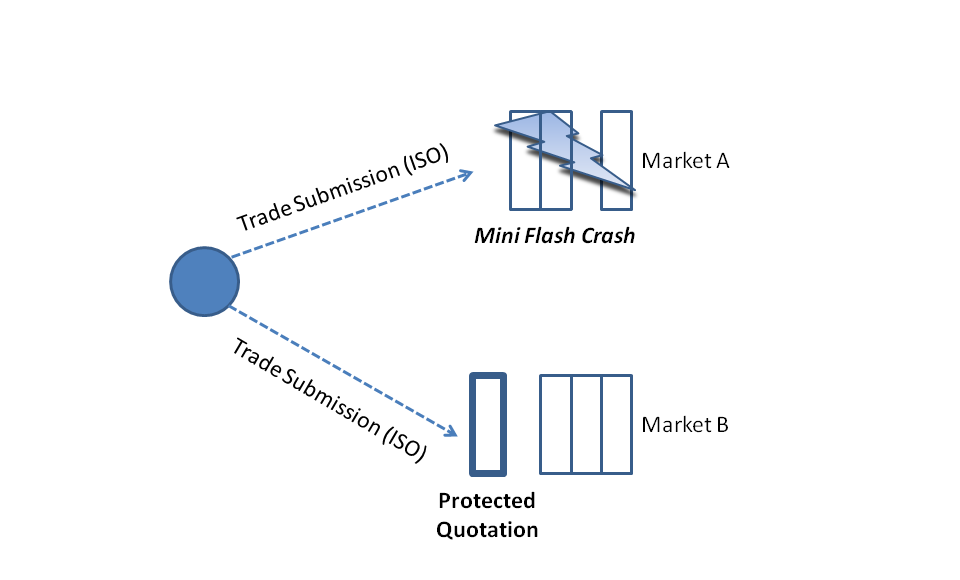}
        \caption{\footnotesize \textbf{ISO Initiated Mini Flash Crash}. A trader submits a package of ISO-marked orders to exchange A and simultaneously sends ISO-marked orders matching the size of Protected Quotes to all trading venues quoting at the NBBO. ISO-marked orders rip through the order book at exchange A resulting in a Mini Flash Crash, despite the additional liquidity available deeper in the order book at other trading venues.}
    \end{center}
\end{figure}
\newpage

In practical terms, allowing the first $k$ trades to be marked as non-ISO, manifests in the dataset as the execution of several trades within the best quotation, either marked as ISO or non-ISO, followed by a series of ISO-marked trades that sweep the order book at a single exchange, executed outside of the best quotation. \\

The second distinct type of Mini Flash Crashes, although less common than the ISO-initiated Mini Flash Crashes, have trades marked as regular. Prior to this type of crashes occurring, we note that all trading venues quoting at the NBBO have their Top of the Book cleared with trades preceding the crashes. Once the Protected Quotation is matched with trades, the subsequent trades constituting the crash occur on a single exchange. It is straightforward to explain what happens in these crashes\footnote{Dennis Dick of Bright Trading was the first to provide an example of such a Mini Flash Crash and explain what occurs - see (Dick, 2011)}: when an exchange receives an order and it is not displaying the NBBO, in order to comply with the Order Protection rule, it has to route part of the order to exchanges showing the NBBO. After the Protected Quotation on other exchanges has been cleared, the exchange is free to fill the remaining part of the order, regardless of whether there is a better quotation deeper in the order book at other trading centres. Figure 4 illustrates this procedure.\\

\begin{figure}[H]
    \begin{center}
        \includegraphics[scale=0.85]{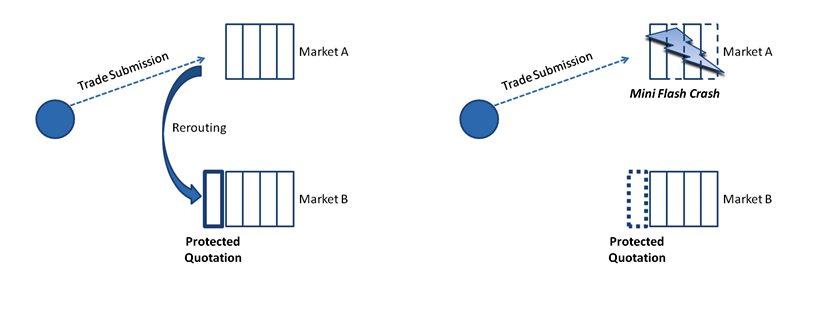}
        \caption{\footnotesize \textbf{Auto-routing initiated Mini Flash Crash}. A trader submits an order to exchange A and, as exchange A is not displaying the NBBO, the part of the order matching the size of the Protected Quotation at exchange B will be routed to comply with the Order Protection rule. Once the Protected Quotation at exchange B is cleared, the resulting balance of orders will be filled at exchange A regardless of the prevailing liquidity at other trading venues, resulting in a Mini Flash Crash.}
    \end{center}
\end{figure}

Based on this example, we define a second type of Mini Flash Crash. In order for a crash to be classified as \textbf{\textit{auto-routing-initiated}}, it has to satisfy the following:
\begin{enumerate}[(i)]
	\item the conditions defining a Mini Flash Crash set out in Section 2 have to be satisfied,
	\item trades constituting the Mini Flash Crash have to be marked as regular, except for the first $k$ trades that can be marked as non-regular if they are executed with the least aggressive market-wide available best quotes in the previous one (1) second,
	\item prior to execution of regular marked trades constituting the Mini Flash Crash, the Top of the Book has to be cleared for all exchanges quoting at the NBBO.
\end{enumerate}

The justification for allowing the first $k$ trades to be marked as non-regular is similar to that in the ISO-initiated Mini Flash Crash definition. The last condition requires that quotes matching the NBBO are cleared as a part of the routing procedure of trades constituting the auto-routing-initiated Mini Flash Crash.\\

Following these definitions we classify 3488 (67.85\%) of crashes as ISO-initiated and 238 (4.64\%) as auto-routing-initiate; the remaining 1414 crashes (27.,51\%) are left unclassified.While the two types of Mini Flash Crashes  are different in their underlying mechanics, both types are the result of the lack of protection for Depth of Book quotations. In many cases there is a significant amount of liquidity on the other exchanges during the time frame of interest, available deeper in the order book. However, as the Order Protection rule only protects the Top of the Book, trades can occur at vastly inferior prices resulting in Mini Flash Crashes in a particular exchange.\\

Finally, we speculate on {\it who} might be causing Mini Flash Crashes. As stressed earlier, most crashes happen due to aggressive use of ISOs. ISO-initiated trades can only be executed by broker-dealers and traders with sponsored access. As typical institutional or retail investors do not have access to such trading mechanisms, they can clearly be ruled out as potential perpetrators. Given the speed and the magnitude of the crashes, it appears likely that Mini Flash Crashes are caused by HFT activity.\\

\section{Liquidity Shocks}
\subsection{Previous Studies}
Despite the criticisms of computerized trading by the popular press and market participants, academic work has found little evidence that the practice is detrimental to financial markets. Recent studies show that, in general, computerized trading improves traditional measures of market quality and contributes to price discovery. Hendershott and Rioran (2009) study the 30 largest DAX stocks on the Deutche Boerse and find that Algorithmic Trading (AT) represents a large fraction of the order flow and contribute more to price discovery than their human counterparts. Moreover, algorithmic traders are more likely to be at the inside quote when spreads are high than when spreads are low suggesting that algorithmic traders supply liquidity when it is expensive and demand liquidity when it is cheap. The authors find no evidence that AT increases volatility. Hendershott, Jones and Menkveld (2011) examine the impact AT has on market quality of NYSE listed stocks. Using normalized measure of NYSE message traffic surrounding the NYSE's implementation of automatic quote dissemination in 2003, they find AT narrows spreads, reduces adverse selection, and increases the informativeness of quotes, especially for larger stocks. Hasbrouck and Saar (2012) measure HFT activity by identifying "strategic runs" of submission, cancellation, and executions in NASDAQ order book. The authors find that HFT improves market quality through decreasing short term volatility, spreads and depth of the order book. Menkveld (2011) claims a large high frequency trader provides liquidity and its entrance corresponds with the decrease in spreads. Brogaard (2010) examines the impact of HFT on the US equity market using a unique HFT dataset for 120 stocks listed on NASDAQ. Brogaard finds that HFT adds to price discover, provides the best bid and offer quotes for a significant portion of the trading day, and reduces volatility. However the extent to which HFT improves liquidity is mixed as the depth high frequency traders provide to the order book is a quarter of that provided by non-high frequency traders.\\

Although the previous studies applied standard approaches of econometrics, they are bounded by strong assumptions and only provide a summary parameters typical of econometric analysis. There is a need for forensic studies, taking into account different assumptions and using different techniques to search and critically examine the data. Although these studies have looked at the fine grain level of time in which algorithms trade, they have focused on the great mass of the data where individual events such as Mini Flash Crashes, though extreme and unusual, can be easily overlooked. The standard econometric tests will be insensitive to Mini Flash Crashes, because these are minor in the context of the length of these incidents and frequencies. In this section we examine market quality by placing Mini Flash Crashes at the centre of our analysis with an event study and evaluate the difference in market liquidity 60 seconds before and after the occurrence of Mini Flash Crashes. Contrary to previous studies, we find strong evidence of HFT having an adverse impact on liquidity during Mini Flash Crashes.\\

\subsection{Spread Difference}

Firstly, we evaluate the difference in spread before and after the occurrence of Mini Flash Crashes. We define midprice as the average of the quoted bid and offer
\begin{equation}
	MidPrice = \frac{offer+bid}{2}
\end{equation}
and compute the spread as:
\begin{equation}
	Spread = \frac{offer-bid}{MidPrice}\cdot 100\%.
\end{equation}
The quoted spread is divided by midprice in order to account for pricing differential between low priced and high priced stocks.
We evaluate the midprice spread resulting from the quotes of the exchange where the Mini Flash Crash will occur, termed Exchange Spread, and also observing the NBBO derived midprice spread, termed NBBO Spread. For the NBBO Spread the offer and bid quote are the  National Best Offer and Bid disseminated by the CQS. If the NBBO quotes are locked or crossed we assign NBBO Spread of 0. The data is smoothed over 100 millisecond time interval. For presentation in Figure 5. the midprice is scaled to zero in the Y-axis.\\

\begin{figure}[h]
    \begin{center}
        \includegraphics[scale=0.65]{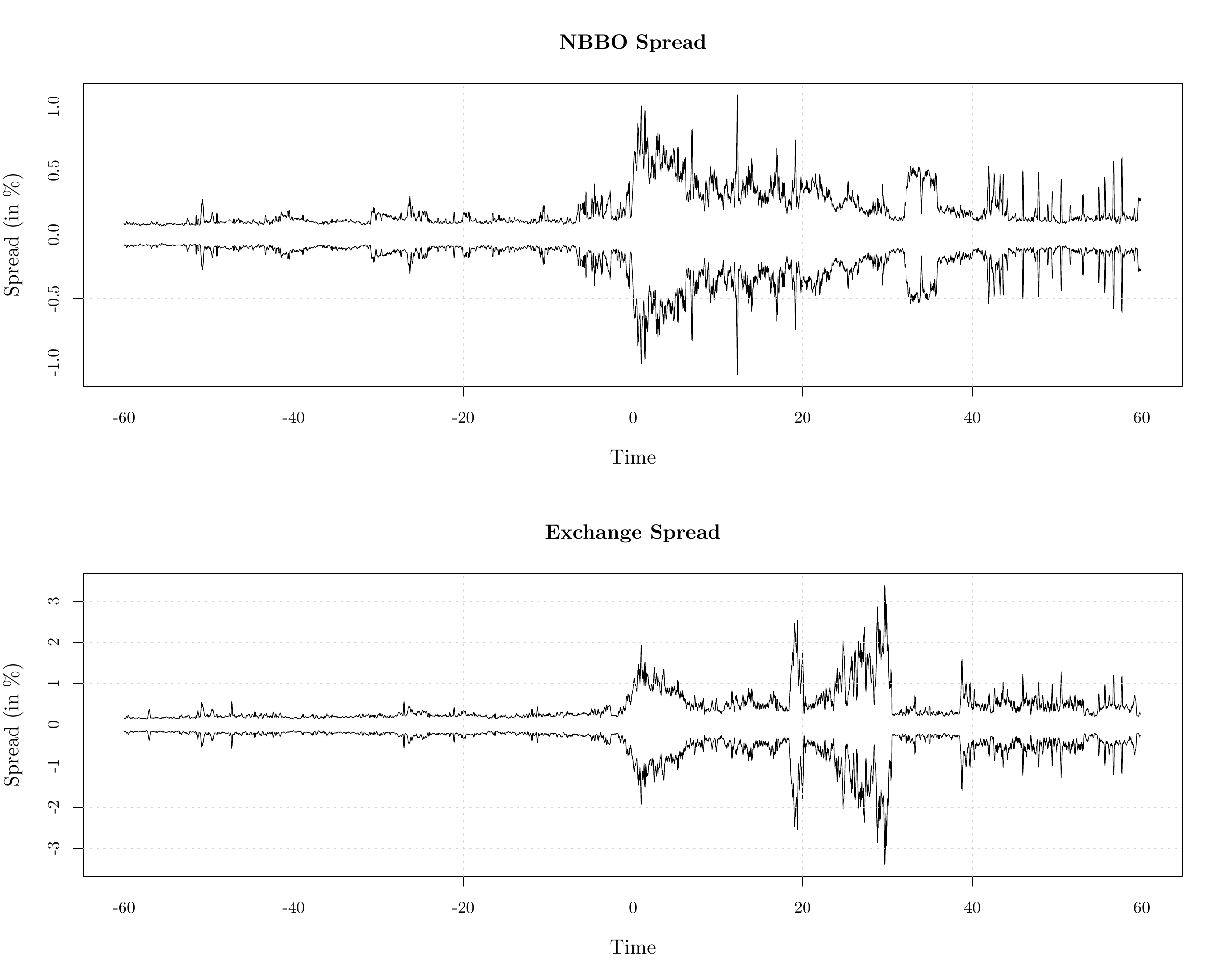}
        \caption{\footnotesize \textbf{NBBO Spread and Exchange Spread.} The figure shows a NBBO Spread (upper graph) and Exchange Spread (lower graph) in the period of 60 seconds before and after the occurrence of a Mini Flash Crash. The midprice is scaled to 0 on the Y-axis.}
    \end{center}
\end{figure}

Figure 5 shows that Mini Flash Crashes have detrimental impact on the Exchange Spread and NBBO Spread. In the 60 seconds prior to the occurrence of a Mini Flash Crash both NBBO Spread and Exchange Spread are stable and equal on average to $0.24\%$ and $0.46\%$ respectively. Both NBBO Spread and Exchange Spread rapidly increase after the Mini Flash Crash to $0.58\%$ and $1.32\%$ in the following 60 seconds, which amounts to increases of $141.60\%$ and $188.05\%$ respectively. Furthermore, we notice that even though Mini Flash Crashes occur extremely fast (within 1.5 seconds by definition), their impact on quoted spread persisted for the whole 60 seconds afterwards. \\

\subsection{Locked and Crossed NBBO Quotes}
Next, we evaluate the percentage of locked and crossed NBBO quotes among all quotes before and after the Mini Flash Crashes. Locked and crossed NBBO quotes occur in a given security when the National Best Bid is higher than, or equal to, the National Best Offer. Rule 611 of Regulation NMS prohibits submission of a quote that would lock or cross the market. Crossed NBBO quotes present risk-less profit for the fastest traders as they can profit by simultaneously buying the security at the crossed offer quote and selling it at the crossed bid quote. Figure 6 shows the results of the analysis, 60 seconds prior and after the Mini Flash Crash.
\begin{figure}[h]
    \begin{center}
        \includegraphics[width=16cm, height=14cm]{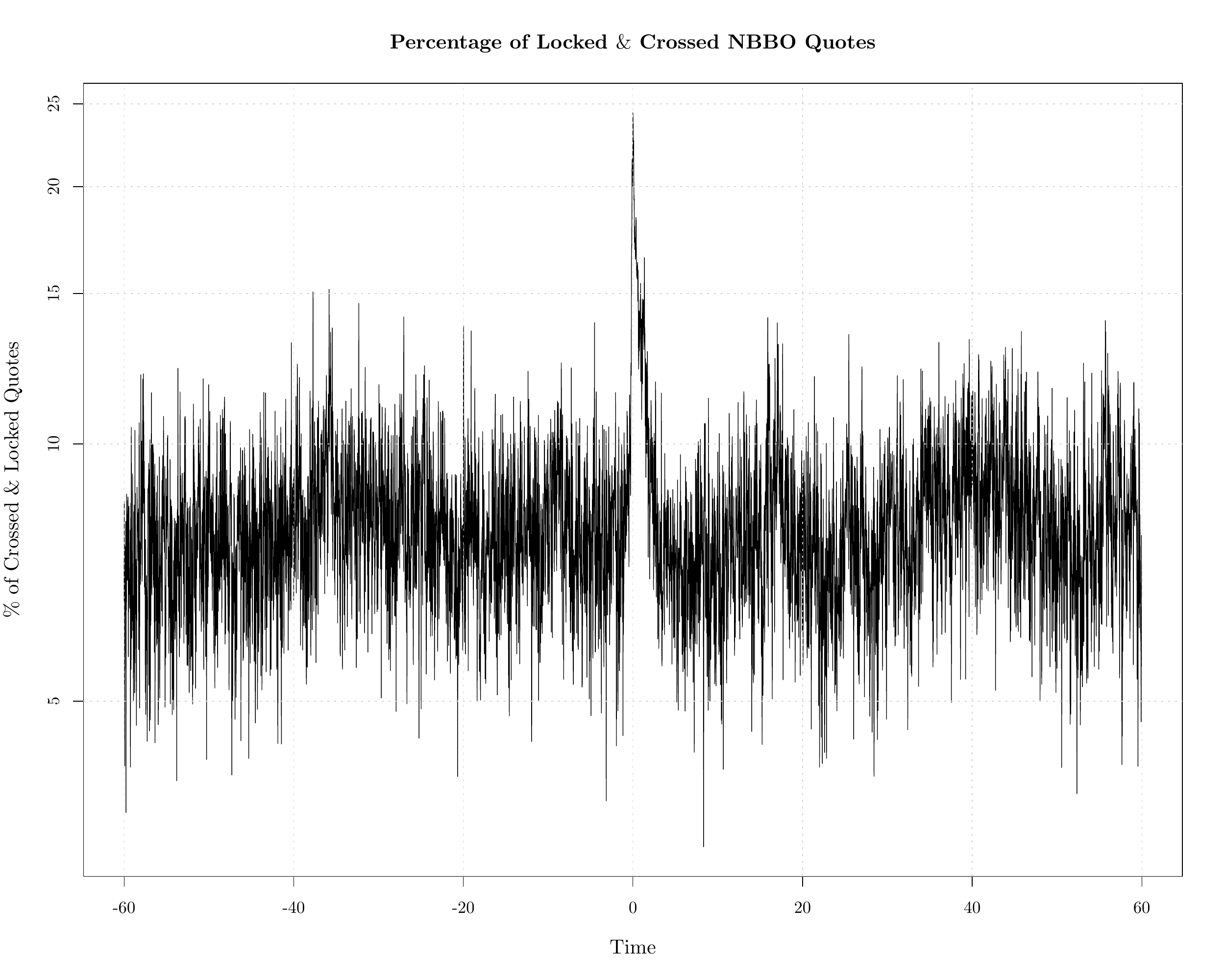}
        \caption{\footnotesize \textbf{Locked $\&$ Crossed NBBO Quotes.} The figure shows the percentage of locked and crossed NBBO quotes in the 60 seconds period prior to the occurrence of a Mini Flash Crash, and the 60 seconds period after the occurrence of a Mini Flash Crash.}
    \end{center}
\end{figure}
Mini Flash Crashes have major impact on the percentage of locked or crossed NBBO quotes, which averaged $8.01\%$ in 60 seconds before the crash, jumps to $24.40\%$ after to crash and declines to $8.23\%$ in the 60 after the crash.

\newpage
\subsection{Quoted Volume}
Finally we analyse the average quoted volume, as measured by the number of shares, at the NBBO before and after the Mini Flash Crashes. Figure 7 presents our results.\\

\begin{figure}[h]
    \begin{center}
        \includegraphics[scale=0.65]{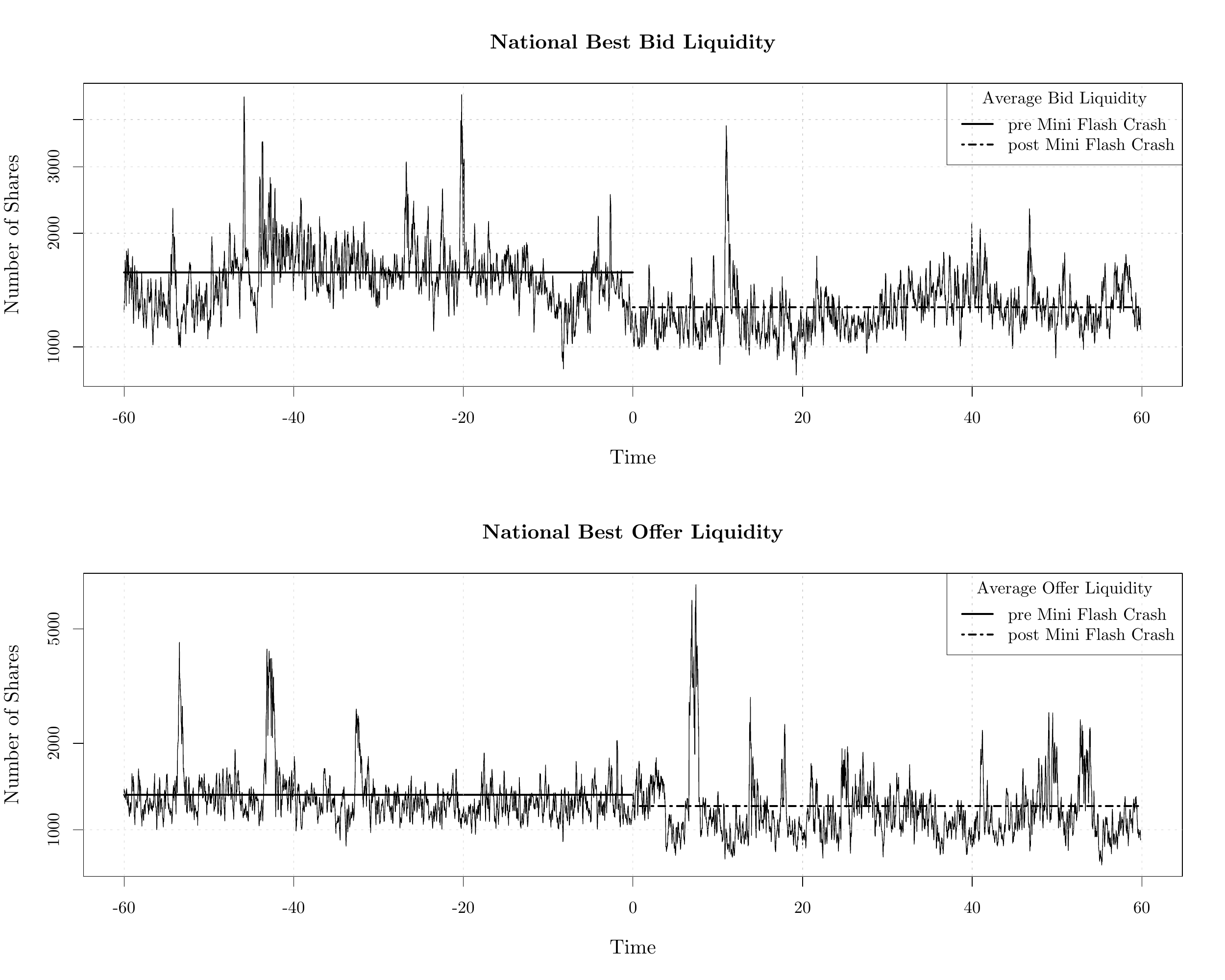}
        \caption{\footnotesize \textbf{Quoted Liquidity at NBBO.} The figure shows the average quoted liquidity at the National Best Bid (upper graph) and National Best Offer (lower graph) in the 60 second period before and after the occurrence of s Mini Flash Crash. The Y-axis is in log scale.}
    \end{center}
\end{figure}

It shows that quoted volume at the NBBO decreases after the Mini Flash Crashes. Before the crash the average quoted volume at the National Best Bid and Offer are, respectively 1575.05 and 1323.49 shares, which drop to 1272.05 shares (reduction of $19.23\%$) and  1210.30 (reduction of $9.35\%$) after the crash. \\

For the exchanges where the Mini Flash Crash occurred, the statistics are 1196.9 and 991.58 shares, respectively for best Bid and Offer, which drops to 870.42 (reduction of $37.51\%$) and 990.99 shares (reduction of $0.06\%$). Figure 8 illustrates these results.\\

\begin{figure}[h]
    \begin{center}
        \includegraphics[scale=0.65]{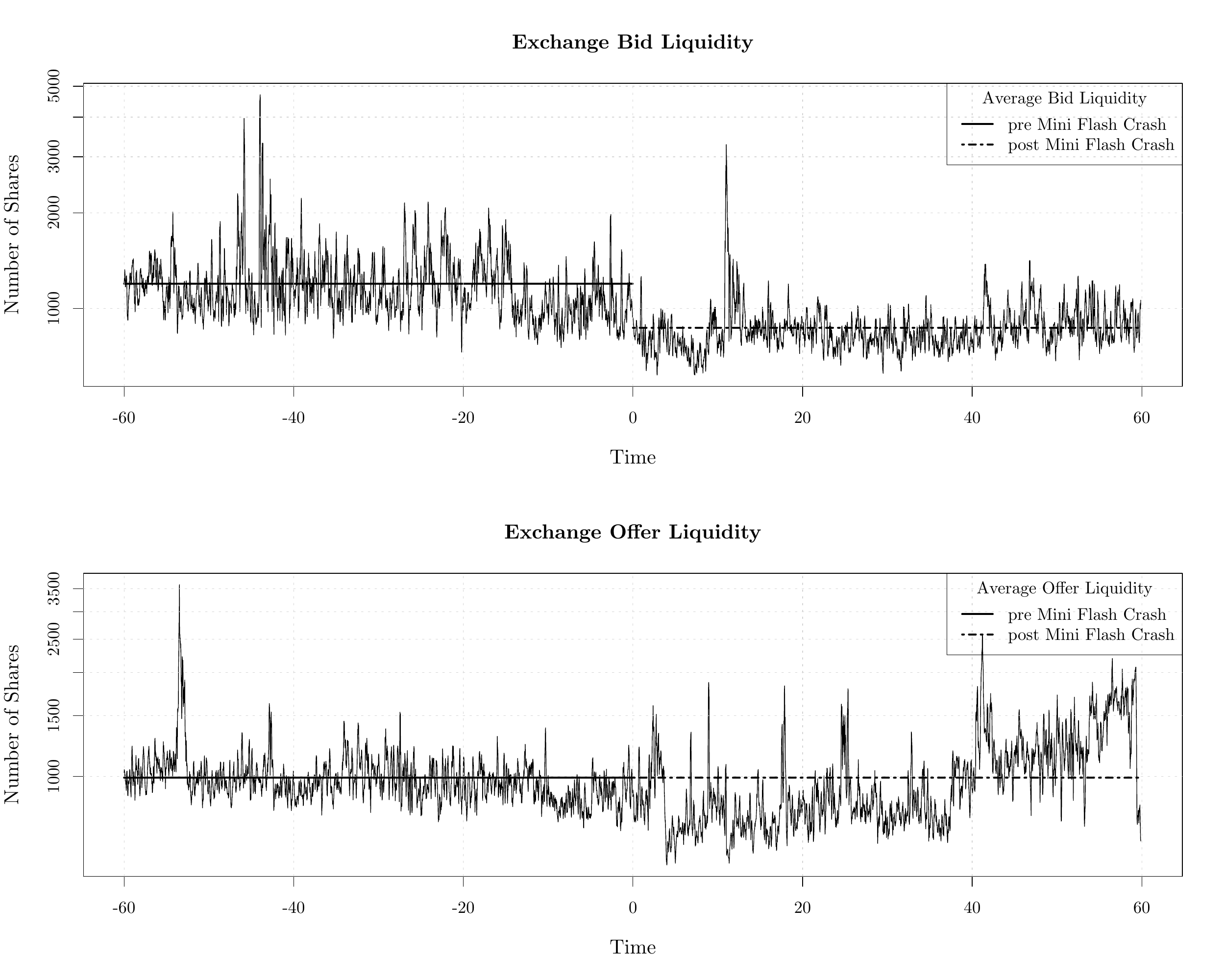}
        \caption{\footnotesize \textbf{Quoted Liquidity at Exchange where Mini Flash Crashes occur.} The figure shows the quoted liquidity at the exchange Best Bid (upper graph) and exchange Best Offer (lower graph) in the 60 second period before and after the occurrence of Mini Flash Crash}
    \end{center}
\end{figure}

From the analysis of quoted liquidity we note that even though there is reduction in quoted liquidity both on the offer and bid side of the market after the occurrence of Mini Flash Crashes, this reduction is asymmetric as the decrease is greater on the bid, and in fact there appeared to be sell-side pressure in the markets after the crash.
\section{Fleeting Liquidity}

Today's traders operate in a tiered market environment where co-location and direct data feeds have made it possible for some participants to obtain tremendous speed advantage, rapidly submitting orders and instantaneously cancelling them. For instance, Hasbrouck and Saar (2012) note that traders (e.g. broker-dealers, market makers or HF traders) can respond to changes in quotes on NASDAQ limit order book in 2 to 3 milliseconds. The speed in which the quotes are posted and cancelled has been criticised by market participants as it creates a false sense of overpriced supply and demand for a stock. Sean Hendelman, CEO at T3 Capital, articulated this concern, ``People are relying on the [stock quote] data and the data is not real.".\footnote{``SEC Probes Canceled Trades'', Lauricella, T. and Strasburg, J. \textit{The Wall Street Journal}, September 2, 2010.} We call such artificial liquidity, Fleeting Liquidity.\\

Many researchers have begun to analyse the implication of this speed advantage. For instance, Cohen and Szpurch (2012) consider the case of a single security market based on a limit order book and two investors, with different speeds of trade execution and show that the faster investor front-runs the slower investor and hence  earns risk free profit. Further, van Kervel (2012) claims that combined liquidity offered on the limit order books may strongly overestimate the actual liquidity available to investors, due to HF traders operating as market makers, duplicating their limit order schedules on several venues to increase their execution probabilities. This is confirmed by empirical results that show trades on the most active venues are followed by cancellations of limit orders on competing venues. McInish and Upson (2012) show empirically that latency differences allow fast liquidity suppliers to pick off slow liquidity demanders at prices inferior to the NBBO. They estimate that fast traders earn more than \$233 million per year at the expense of slow traders. Watson, Van Ness and Van Ness (2012) find a structural break between 2001-2005 and 2005-2010 where there is a dramatic increase in the number of limit orders which are cancelled. They document commonality in the cancellation rates in all stocks of an exchange or across exchanges. \\

Mini Flash Crashes present a unique opportunity to analyze displayed liquidity at a time of severe stress, as it is presumed that due to the speed of the crashes it is not possible to predict them and withdraw liquidity by cancelling resting limit orders. Further, it has been claimed that price moves in the stock market are usually not caused by market makers backing away. Specifically, Gregg Berman, a senior adviser to the director of the SEC’s trading and markets division suggest, ``...these events ... were generally not caused by a sudden withdraw of liquidity by market makers.''\footnote{``Market Structure: What we Know, and What we Need to Know'', Berman, G.E., \textit{12 Annual SIFMA Market Structure Conference}, New York, September 21, 2011}.\\

We say that a Mini Flash Crash experienced Fleeting Liquidity if the quotes disseminated by the SIP are not hit during the occurrence of Mini Flash Crash. This means the best displayed quotation was cancelled faster than the SIP could disseminate the removal of these resting limit orders. Table 2 shows the statistics of Fleeting Liquidity phenomenon. We note that the phenomenon of Fleeting Liquidity is as often as $37.99\%$ of Mini Flash Crashes in our sample.\\

Next, we analyse the connection between Fleeting Liquidity and Mini Flash Crashes with a logit regression on the following variables:
\begin{itemize}
	\item  \textit{Time} is the time of the crash in milliseconds; 
	\item  \textit{\%PriceChange} is the absolute percentage price change in a crash;
	\item  \textit{Exch} an indicator variable equal to 1 if the crash occurred on NYSE, 2 if it occurred on Nasdaq, 3 if it occurred on ARCA, 4 if it occurred on AMEX, 5 if it occurred on BATS and 6 if it occurred on ISE;
	\item  \textit{Vol} is the total volume in the crash;
	\item \textit{UpDown} is an indicator variable equal to 0 if it was a down crash, and 1 if it was an up crash;
	\item \textit{NoTrades} is the number of trades in a crash;
	\item  \textit{Type} is an indicator variable equal to 1 if the crash was ISO-initiated, 2 if it was auto-routing-initiated, and 3 if it is unclassified;
	\item \textit{FleetLiq} is the dependent variable whose value is equal 0 if the crash did not experience the phenomenon of Fleeting Liquidity, and 1 if it did;
\end{itemize}   
The logit regression has the following form;
\begin{equation}
	\operatorname{ln}\bigg(\frac{\mathbb{P}(FleetLiq_i=1)}{1-\mathbb{P}(FleetLiq_i=1)}\bigg)= f(X_i|\alpha)
\end{equation}
where the function $f$ equals 
\begin{eqnarray*}
 \lefteqn{f(X_i|\alpha)=}\\
   && ~~~~~~~~~~~\alpha_0 + \alpha_1\cdot Time_i + \alpha_2\cdot \%PriceChange_i + \alpha_3\cdot Exch_i \\
   && ~~~~~~~~~~~ +\alpha_4\cdot Vol_i + \alpha_5 \cdot UpDown_i + \alpha_6\cdot NbTrades_i + \alpha_6\cdot Type_i
\end{eqnarray*}
for $X_i=(Time_i, \%PriceChange_i, Exch_i, Vol_i, UpDown_i, NbTrades_i, Type_i)$ and  $\alpha = (\alpha_0, \alpha_1, \alpha_2, \alpha_3, \alpha_4, \alpha_5, \alpha_6)$. We label the predicted variable $\widehat{FleetLiq}$ with 1 (Fleeting Liquidity) if the predicted probability 
\begin{equation}
	\widehat{p}_i=\frac{1}{1+e^{-f(X_i|\widehat{\alpha})}}
\end{equation}
is greater than 0.5, otherwise we label it with 0 (no Fleeting Liquidity). In Table 2 we present the results of the logit regression: two variables, \textit{Time} and \textit{Vol}, are statistically significant at the 1\% level, and \textit{NbTrades} is statistically significant at the 5\% level, all three coefficients are negative, implying that Fleeting Liquidity is characteristic of ultra-fast Mini Flash Crashes of small volume and a small number of trades. The fitted logit regression has a classification precision of 54.80\%.

\begin{center}
\begin{table}[H]
	\captionsetup{singlelinecheck=off}
	\footnotesize
	\caption[try out1]{\footnotesize \textbf{Statistics of Mini Flash Crashes with Fleeting Liquidity phenomenon and logistic regression results}. \\ 
	This table shows the results of logistic regression; the variables are defined as in the text above.}
	\begin{center}
	\begin{tabular}{l l l l l l}
		& Sep 2008 & Oct 2008 & Nov 2008 & May 2010 & Total \\ \hline
	Total Crashes & 1058 & 2919	& 672 & 	491 & 5140 \\
	~~~~Up Crashes & 527	& 1709 &	 337 & 187 & 2760 \\
	~~~~Down Crashes & 531 & 1210 & 335 & 304 & 2380 \\ 
	& & & & & \\
	Fleeting Liquidity & 377 (35,63\%) & 1114  (38,16\%) & 289 (43,00\%) & 173 (35,23\%) & 1953 (37,99\%) \\ \hline
	& & & & & \\	
	& & & & & FleetLiq \\ \hline
	Intercept & & & & & 0.1444 \\
	& & & & & (1,4678) \\
	Time & & & & & $-0.2065^{**}$ \\
	& & & & & (-2,7481) \\
	\%PriceChange & & & & & 0.1098 \\
	& & & & & (0,7346) \\
	Exch & & & & & 0.0102 \\
	& & & & & (0,2474) \\
	Vol & & & & & -4.6876E-$6^{**}$ \\
	& & & & & (-2,7527) \\
	UpDown & & & & & 0.0022 \\
	& & & & & (0,0738) \\
	NoTrades & & & & & $-0.0067^{*}$ \\
	& & & & & (-2,2836) \\
	Type & & & & & -0.0464 \\
	& & & & & (-1,0201) \\ \hline
	Classification Precision & & & & & 54,80\% \\ \hline
	\multicolumn{3}{l}{$^{**}$Statistically Significant at 1\% Level} & & & \\
	\multicolumn{3}{l}{$^{*}$Statistically Significant at 5\% Level} & & &  \\ \hline
	\end{tabular}
	\end{center}
\end{table}
\end{center}
\section{Discussion}
At the heart of the Regulation NMS is the Order Protection Rule aimed to unify US equity markets and establish intermarket protection against inferior execution. While the regulation has served its purpose initially, the current market environment might not be suitable for an ecosystem dominated by speed traders. Mini Flash Crashes have become the consequence of the fragmented liquidity that is indicative of the current market structure with its myriad of trading venues. As more and more exchanges are created, traders need to be increasingly aware of the current market structure. Unfortunately, the trading in US equity markets is split between 13 public exchanges, more than 30 dark pools and over 200 internalizing broker-dealers (Shapiro, 2010) making it difficult for all participants of the capital markets to access liquidity in an organized manner. As the number of trading venues increases, imbalances in liquidity across these venues will naturally result, creating an opportunity for the fastest traders to profit by scouring the markets for imbalances and eliminating them. Mini Flash Crashes demonstrate that market participants require a better understanding of how their orders take liquidity. The visible liquidity at the time an order was placed does not necessarily remain when the order is executed – but rather the converse.\\

Extending the Order Protection rule to the Depth of the Book may be a quick fix to the Mini Flash Crashes studied here. Traders would then be unable to trade through resting limit orders with superior prices deeper in the order book at other trading venues. However, such a requirement for market participants to comply with the same trade-through protection for the entire Depth of the Book quotations would be extremely costly and complex. In a market environment where fleeting orders located deeper in the order book have been found to contribute little to the overall liquidity (Gai, Yao, Ye, 2012), it is practically impossible to uphold such regulation. Moreover, there is no guarantee that protecting quotations at multiple price levels would encourage the display of limit orders and improve overall liquidity. Further, several unintended consequences may arise from this full order-book protection proposal, as it would create a tremendous amount of confusion as to how and where to route an order. With Fleeting Liquidity, displayed limit orders deeper in the order book might be cancelled faster than it would take an exchange to route an order to the trading venue in question, resulting in partial or no execution of trades.\\

Nonetheless, the aggressive use of ISOs resulting in Mini Flash Crashes should not be tolerated. Note that the ISO package of orders is submitted with a limit price, meaning that traders submitting such orders are well aware that the corresponding execution on trading venues with poor liquidity might result in trades that are well off the NBBO. We suggest that regulators pay particular attention to ISO-initiated Mini Flash Crashes and take appropriate actions against market participants responsible for such Mini Flash Crashes. Finally, we note that the abuse of ISOs has also been mentioned in the context of \textit{latency arbitrage}. Narang (2010) pointed out that U.S. broker-dealer HF traders with access to direct data feeds can profit from the latency of SIP consolidating the NBBO system by using ISO. ISOs allow these market participants to send orders directly to market places while the markets appear to be locked due to the latency of the respective consolidated data feeds. This situation may lead to a violation of price-time priority, as ISOs may directly be posted and executed, while normal orders could not be directly posted and executed because they would appear to lock the market.\footnote{For details of please refer to (Narang, 2010).} In view of this, we also suggest regulators should consider the banning of ISOs.\\
\section{Conclusion}
In this article we analyse Mini Flash Crashes in US equity markets. We find that Mini Flash Crashes are the result of regulatory framework and market fragmentation; in particular due to aggressive use of Intermarket Sweep Orders and Regulation NMS protecting only the Top of the Book. We find that Mini Flash Crashes have an adverse impact on market liquidity resulting in wider spread, increased number of locked and crossed NBBO quotes and decrease in quoted volume. Finally, we document the phenomenon of Fleeting Liquidity associated with ultra-fast crashes Mini Flash Crashes of small volume, and a small number of trades.

\section{Bibliography}
\indent{~~~~~Biais, B., Foucault, T., Moinas, S., (2011), \textit{Equilibrium High Frequency Trading}}\newline
\indent{Brogaard, J. A., (2010),\textit{High Frequency Trading and its Impact on Market Quality}, Working Paper}\\
\indent{Chakravarty, S., Jain, P., Upson, J., Wood, R., (2010), \textit{Clean Sweep: Informed Trading through Intermarket Sweep Orders},  forthcoming, Journal of Financial and Quantitative Analysis}\\
\indent{Cespa, G., Foucault, T., (2012), \textit{Illiquidity Contagion and Liquidity Crashes}}\\
\indent{Cohen, S., N., Szpruch, L., (2012), \textit{A Limit Order Book for Latency Arbitrage}}\\
\indent{Dick, D., (2011), \textit{Unlocking the Mystery Behind the IBM "Flash Dash"}, Bright Trading, http://www.defendtrading.com/blog.html?entry=unlocking-the-mystery-behind-the}\\
\indent{Easley, D., Lopez de Prado, M., O'Hara, M., (2011), \textit{The Microstructure of "Flash Crash": Flow Toxicity, Liquidity Crashes and the Probability of Informed Trading}, The Journal of Portfolio Management, Vol. 32, No. 2, pp. 118-128}\\
\indent{Egginton, J., Van Ness, B., Van Ness, R., (2012), \textit{Quote Stuffing}}\\
\indent{Gai, J., Yao, C., Ye, M., (2012), \textit{The Externality of High Frequency Trading}}\\
\indent{Hasbrouck, J., Saar, G., (2012), \textit{Low-Latency Trading}}\\
\indent{Hendershoot, T., Johnes, M. C., Menkveld A., (2011), \textit{Does Algorithmic Trading Improve Liquidity?}, The Journal of Finance 66, 1-33}\\
\indent{Hendershoot, T., Riordan, R., (2009), \textit{Algorithmic Trading and Information}, Working Paper, NET Institute}\\
\indent{Jarrow, R., A., Protter, P., (2011), \textit{A Dysfunctional Role of High Frequency Trading in Electronic Markets}}\\
\indent{Johnson, N., Zhao, G., Hunsader, E., Meng, J., Ravindar A., Carran, S., Tivnan, B., (2012), \textit{Financial Black Swans Driven by Ultrafast Machine Ecology}}\\
\indent{Kirilenko, A., Kyle, A., S., Samadi, M., Tuzun, T., (2011), \textit{The Flash Crash: The Impact of High Frequency Trading on an Electronic Market}}\\
\indent{McInish, T., Upson, J., (2012), \textit{Strategic Liquidity Supply in a Market with Fast and Slow Traders}}\\
\indent{Menkveld, A. J., (2011), \textit{High Frequency Trading and the New Market Makers}, Working Paper}\\
\indent{Nanex, (2011), \textit{Flash Equity Failures in 2006, 2007, 2008, 2009, 2010, and 2011}, 
http://www.nanex.net/FlashCrashEquities/FlashCrashAnalysis$_{}$Equities.html}\\
\indent{Nanex, (2010), \textit{Nanex Flash Crash Summary Report}, \newline
http://www.nanex.net/FlashCrashFinal/FlashCrashSummary.html}\\
\indent{Nanex, (2010), \textit{May 6'th 2010 Flash Crash Analysis - Final Conclusion}, \newline http://www.nanex.net/FlashCrashFinal/FlashCrashAnalysis$_{}$Theory.html}\\
\indent{Narang, M., (2010), \textit{Tradeworx Presentation on High Frequency Trading}, Tradeworx}\\
\indent{Paddrik, M., Hayes, R., Todd, A., Yang, S., Beiling, P., Scherer, W., (2011), \textit{An Agent Based Model of the E-Mini S{\&}P 500}}\\
\indent{Schapiro, L. M., (2010), \textit{Strengthening Our Equity Market Structure}, US Securities and Exchange Commission Economic Club of New York, New York}\\
\indent{US Securities and Exchange Commission, (2010), \textit{Findings Regarding the Market Events of May 6, 2010}}\\
\indent{van Kervel, V., (2012), \textit{Liquidity: What You See is What You Get?}}\\
\indent{Watson, E., Van Ness, B., Van Ness, R., (2012), \textit{Cancelling Liquidity}} \\
\indent{Wood, R., Upson, J., McInish, T., H., (2012), \textit{The Flash Crash: Trading Aggressiveness, Liquidity Supply, and the Impact of Intermarket Sweep Orders}}\\

\end{document}